\documentclass[sigconf]{acmart}
\AtBeginDocument{%
  }

\usepackage{algorithm}
\usepackage{amsmath}
\usepackage{subcaption}
\usepackage[noend]{algpseudocode}
\usepackage{algorithmicx}
\usepackage{colortbl} 
\usepackage{xcolor}
\usepackage{multirow} 
\newcommand{\wyx}[1]{\textcolor{black}{#1}}
\newcommand{\wyxnew}[1]{\textcolor{black}{#1}}

\begin{document}

\title{Adaptive Hybrid Caching for Efficient Text-to-Video Diffusion Model Acceleration}

\author{Yuanxin Wei}
\affiliation{%
  \institution{Sun Yat-sen University}
  \city{Guangzhou}
  \country{China}}

\author{Landsong Diao}
\affiliation{%
  \institution{Alibaba Group}
  \city{Beijing}
  \country{China}}

\author{Bujiao Chen}
\affiliation{%
  \institution{Alibaba Group}
  \city{Beijing}
  \country{China}}

\author{Shenggan Cheng}
\affiliation{%
  \institution{National University of Singapore}
  \country{Singapore}}

\author{Zhengping Qian}
\affiliation{%
  \institution{Alibaba Group}
  \city{Beijing}
  \country{China}}

\author{Wenyuan Yu}
\affiliation{%
  \institution{Alibaba Group}
  \city{Beijing}
  \country{China}}

\author{Nong Xiao}
\affiliation{%
  \institution{Sun Yat-sen University}
  \city{Guangzhou}
  \country{China}}

\author{Jiangsu Du}
\affiliation{%
  \institution{Sun Yat-sen University}
  \city{Guangzhou}
  \country{China}}
  



\begin{abstract}
\wyxnew{Efficient video generation models are increasingly vital for multimedia synthetic content generation.} 
Leveraging the Transformer architecture and the diffusion process, video DiT models have emerged as a dominant approach for high-quality video generation. However, their multi-step iterative denoising process incurs high computational cost and inference latency, which limits their practical deployment in large-scale and interactive multimedia applications.
Caching, a widely adopted optimization method in DiT models, leverages the redundancy in the diffusion process to skip computations in different granularities (e.g., step, cfg, block). Nevertheless, existing caching methods are limited to single-granularity strategies, struggling to balance generation quality and inference speed in a flexible manner. In this work, we propose MixCache, a training-free caching-based framework for efficient video DiT inference. MixCache first distinguishes the interference and boundary between different caching strategies, and then introduces a context-aware cache triggering strategy to determine when caching should be enabled, along with an adaptive hybrid cache decision strategy for dynamically selecting the optimal caching granularity. Extensive experiments on diverse models demonstrate that MixCache can significantly accelerate video generation (e.g., 1.94× speedup on Wan 14B, 1.97× speedup on HunyuanVideo) while delivering both superior generation quality and inference efficiency compared to baseline methods.
\end{abstract}




\received{20 February 2007}
\received[revised]{12 March 2009}
\received[accepted]{5 June 2009}

\maketitle
 
\section{Introduction}
Diffusion Transformer (DiT)~\cite{DiT} has revolutionized video generation by integrating the scalability of the Transformer architecture~\cite{transformer} with the power of the diffusion process~\cite{DDPM, Diffusion-2}, enabling unprecedented quality in synthetic video creation.
Recent advances in video DiT models, such as SD3.0~\cite{SD3.0}, Sora~\cite{Sora}, CogVideoX~\cite{CogvideoX}, and Wan~\cite{Wan}, have greatly expanded the capabilities of multimedia systems, supporting a wide range of applications including text-to-video generation~\cite{Text2Video-Zero,Tune-A-Video}, video editing~\cite{VideoComposer,VACE}, and video continuation~\cite{CogvideoX}.
\wyxnew{Such advancements not only improve the fidelity and variety of generated multimedia content but also open new possibilities for creative expression in video-based applications.}

Despite achieving superior fidelity, the inference of video DiT models relies on an iterative denoising process, which demands substantial computation and hinders time-sensitive deployment.
Starting from a random Gaussian noise initialization, these models require tens of denoising steps, typically ranging from 20 to 100, to progressively reconstruct high-quality video.
Consequently, generating a 5-second 720p video with a single GPU can take 50 minutes, presenting a significant latency bottleneck.

\begin{figure}[t]
\centering
\includegraphics[width=1.0\columnwidth]{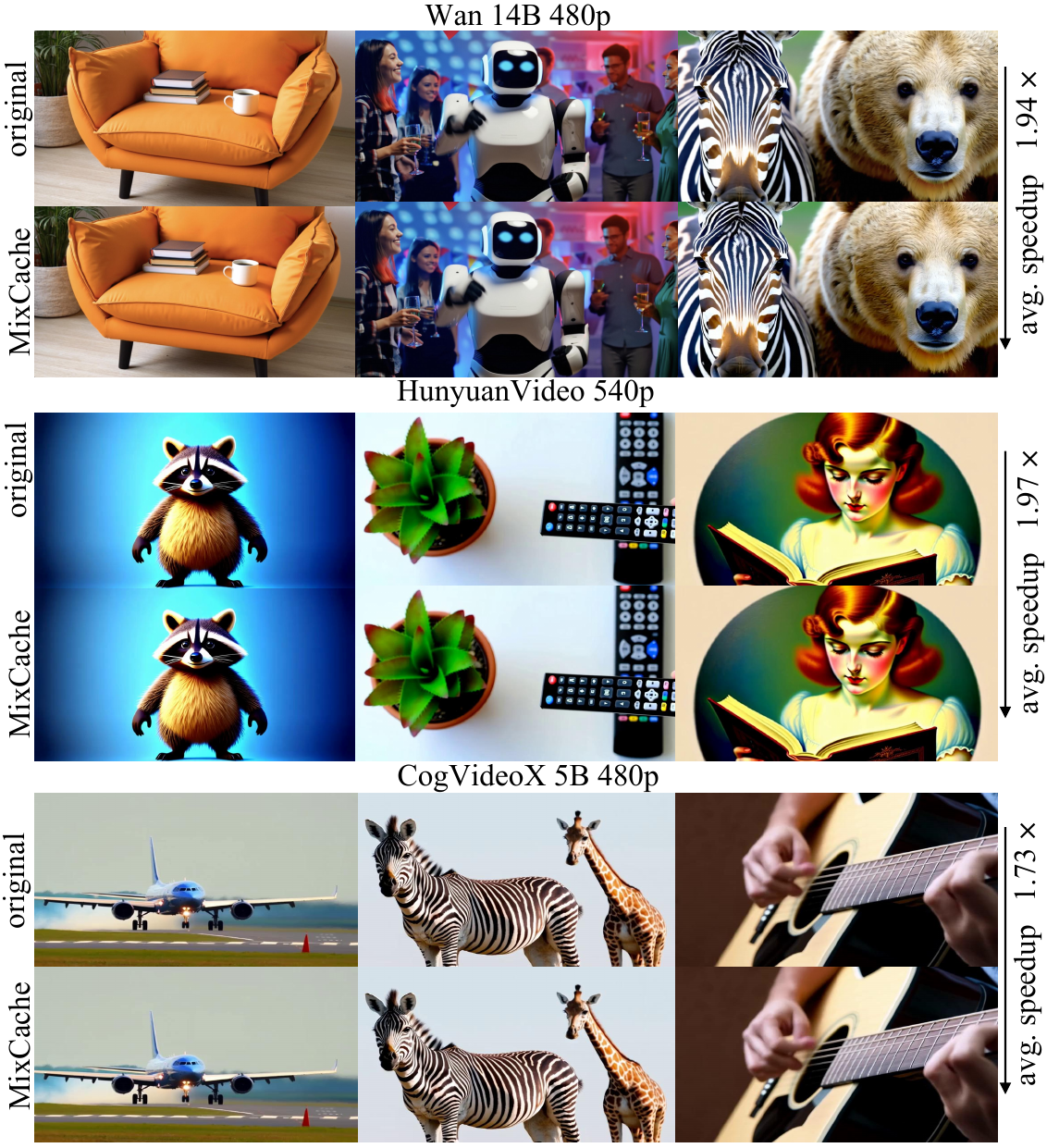}
\caption{MixCache visualization across video DiT models. }
\label{fig:vis_page_1}
\end{figure}

The research community has made significant efforts to speed up video DiT models.
Among them, caching-based acceleration has emerged as one of the most widely adopted approaches.
Caching can reduce redundant computation and improve efficiency by storing and reusing intermediate features, leveraging output similarity across diffusion timesteps.
According to the caching granularity from coarse to fine, existing caching methods can be divided into step level~\cite{TeaCache,AdaCache,MagCache}, cfg level~\cite{FasterCache}, block level~\cite{DeepCache,Block-Caching,FasterDiffusion,FORA,PAB,L2C,MD-DiT}, and token level~\cite{TokenCache,DuCa,ToCa}.

However, existing caching methods primarily exploit output redundancy at a single granularity, overlooking the inherent redundancy across multiple granularities throughout the diffusion process.
This narrow focus compromises the balance between generation quality and computational efficiency. 
Through systematic investigation, we demonstrate that adaptively combining caching methods at diverse granularities enables more effective utilization of output redundancy. 
In this paper, we propose MixCache, a training-free, caching-based inference framework for video DiT models that enables flexible integration of caching methods with varying granularities.
This hybrid caching framework provides more effective acceleration for video DiT inference while preserving generation quality.

Our contributions are as follows:
\begin{itemize}
    \item We conduct a comprehensive analysis of redundancy across multiple granularities in the diffusion process, including step level, cfg level and block level, and reveal the dynamism nature of redundancy.
    \item We propose a context-aware cache triggering strategy to determine when to enable caching, along with an adaptive hybrid cache decision strategy to determine the caching granularity (step/cfg/block) in a flexible manner for each timestep.  
    \item Building upon the above strategies, we present MixCache, a training-free caching-based inference framework that adaptively integrates multi-granularity caching methods without modifications to model structure.
    \item Extensive experiments on industrial-scale video DiT models show that MixCache demonstrates significant improvement in inference speed while maintaining high video quality.
\end{itemize}
\section{Related Works}
In the following, we introduce works that are highly related to our method.

\textbf{Step level caching.}
Recent DiT advancements introduce diverse step level caching strategies.
TeaCache~\cite{TeaCache} leverages input-output correlations by estimating timestep embeddings to implement a dynamic caching. 
AdaCache~\cite{AdaCache} dynamically evaluates step-wise differences to determine skip lengths. 
AB Cache~\cite{AB-Cache} extends prior methods by reusing combined outputs from the previous $k$ steps, rather than a single-step result. 
NIRVANA~\cite{NIRVANA} diverges from these intra-generation approaches by enabling cross-generation cache reuse to skip computation of preceding sampling timesteps.

\textbf{CFG level caching.}
Recent studies leverage cfg similarities to accelerate diffusion inference. 
FasterCache~\cite{FasterCache} exploits the similarity between conditional and unconditional outputs at the same timestep to construct a cfg level cache mechanism. 
DiTFastAttn~\cite{DiTFastAttn} identifies consistent attention patterns in specific attention heads of conditional and unconditional inference, implementing cfg level attention sharing to bypass redundant computation.

\textbf{Block level caching.}
For the DiT architecture, recent works propose block level caching with varying strategies. FORA~\cite{FORA}, PAB~\cite{PAB}, and BlockDance~\cite{BlockDance} employ static interval-based block skipping (e.g., MLP/attention blocks). $\Delta$-DiT~\cite{Delta-DiT} adopts a phase-specific caching strategy, storing back blocks during early stages and front blocks in later stages. L2C~\cite{L2C} and MD-DiT~\cite{MD-DiT} implement dynamic runtime caching through learnable layer selection and gradient-free search mechanisms, respectively. 


\textbf{Other optimizations.}
Recent works~\cite{TokenCache,DuCa,ToCa,FastCache} have explored token redundancy (such as background areas) to reduce computational cost. 
However, the dynamic nature and dataset-dependent characteristics of input tokens necessitate extensive code modifications to transformer blocks for implementation, resulting in poor compatibility. Moreover, the acceleration efficiency of token level caching is fundamentally limited by its fine-grained caching granularity and the overhead of token importance scoring.
In response to the performance bottleneck caused by the multi-step iterative characteristic of the diffusion process, some works also explore efficient solvers, including DDIM~\cite{DDIM}, DPM-Solver~\cite{DPM-Solver} and DPM-Solver++~\cite{DPM-Solver++}.
Additionally, distillation-based methods~\cite{Progressive-Distillation,Knowledge-Distillation,GuidedDistillation} have been developed to compress the number of diffusion sampling timesteps.

\section{Methodology}

\subsection{Preliminary}
\textbf{Diffusion model} is a generative model consisting of a forward process and a reverse process.
In the forward diffusion process, given an original video $x_0$ and a random timestep $t$, the video after $t$ diffusion timesteps is:
\begin{equation}
x_t = \sqrt{\delta_t} x_{t-1} + \sqrt{1 - \delta_t} \epsilon_t, \quad t=1,2,...,T
\label{eq:denoise}
\end{equation}
where $\delta_t$ is a schedule coefficient that varies with $t$, and $T$ is the total sampling timesteps.
A noise estimation network plays a critical role in approximating the noise distribution in the diffusion process. Specifically, it aims to minimize the discrepancy between the predicted noise term $\epsilon_{\theta}$ and the actual noise $\epsilon$. 
In most current works, the noise estimation network adopts the DiT architecture, where the predicted noise function $\epsilon_{\theta}(x_t)$ can be further reformulated as:
\begin{equation}
\begin{aligned}
\epsilon_{\theta}(x_t) &= f_{L-1}(f_{L-2}(\dots(f_0(x_t)))) \\
&= f_{L-1} \circ f_{L-2} \circ \cdots \circ f_0(x_t)
\end{aligned}
\label{eq:DiT}
\end{equation}
where $f_n$ represents the $n$-th DiT block and $L$ represents the total number of DiT blocks.

The inference process, defined as the reverse transformation of noisy data into clean output, is a crucial part of the diffusion process. 
Initially, a random Gaussian noise $X_T$ is given. It is input into the noise estimation network $\epsilon_{\theta}$ to obtain the noise estimate $\epsilon_{\theta}(x_T)$.
According to specific sampling solver $\Phi$, the noisy video is denoised to produce the denoised sample $x_{T-1}$ after one timestep.
After iterating this process $T$ times, the final generated video is obtain.
\begin{equation}
x_{t-1} = \Phi(x_t,t,\epsilon_{\theta}(x_t)), \quad t= T, T-1, ...,1
\label{eq:reverse}
\end{equation}


\textbf{Classifier-Free Guidance (CFG)} has proven to be a powerful technique for improving the fidelity of generated videos in diffusion models. 
During the sampling process, CFG generates two distinct predictions: the conditional output $\epsilon_{\theta}(x_t,t,c)$ conditioned on the input context $c$, and the unconditional output $\epsilon_{\theta}(x_t,t,\phi)$ derived from the empty/negative prompt $\phi$.
The final denoised output is:
\begin{equation}
\tilde{\epsilon}_{\theta}(x_t,t,c) = (1+g)\cdot\epsilon_{\theta}(x_t,t,c)-g\cdot\epsilon_{\theta}(x_t,t,\phi)
\label{eq:cfg}
\end{equation}
where $g$ is the guidance scale. 
While CFG significantly enhances visual quality, it also increases computational cost and inference latency due to the additional computation required for unconditional outputs.

\subsection{Analysis and Motivation}


\begin{figure}[t]
\centering
\includegraphics[width=1.0\columnwidth]{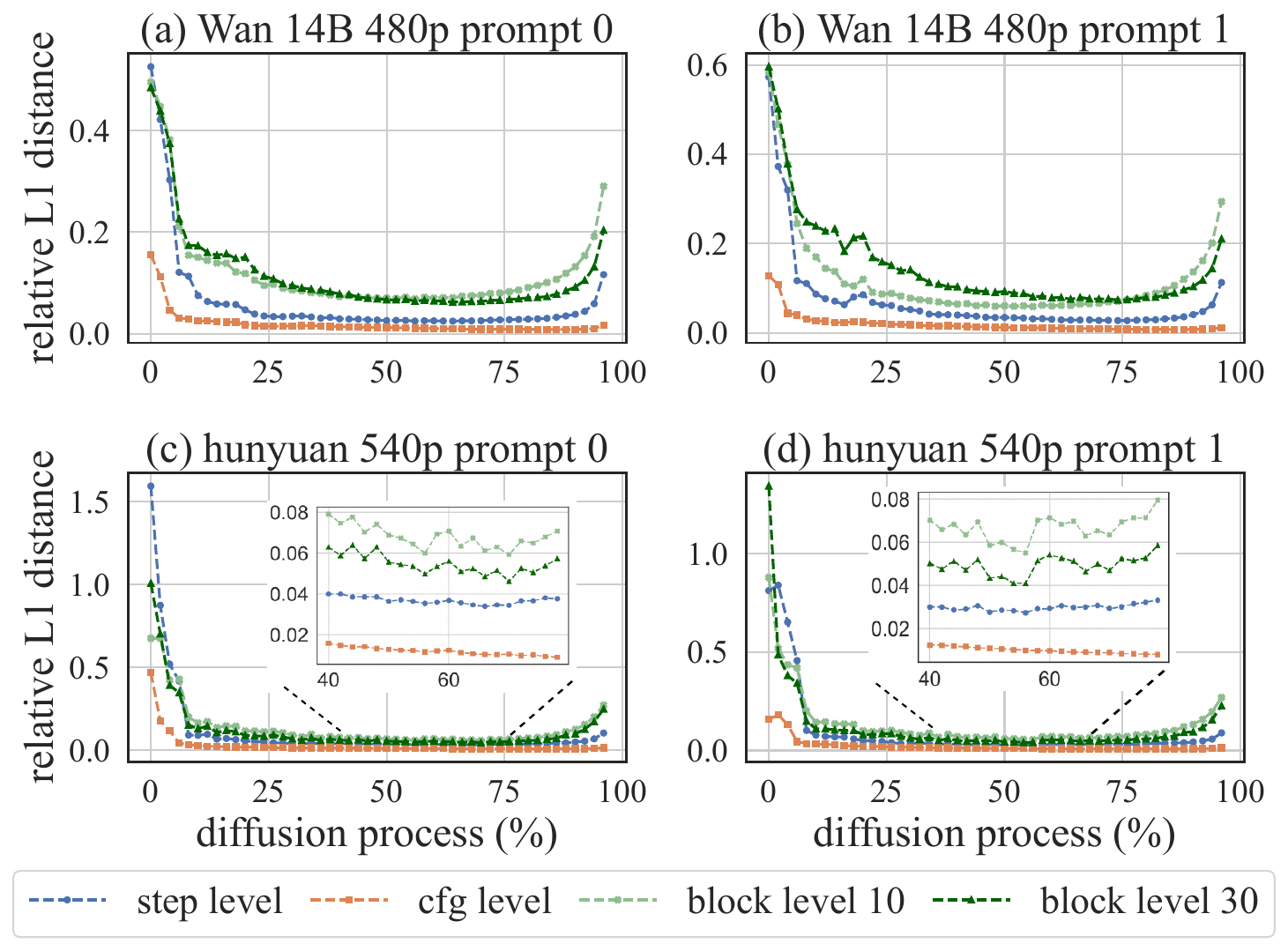}
\caption{Three levels of redundancy across denoising timesteps in Wan 14B 480p and HunyuanVideo 540p.}
\label{fig:diff}
\end{figure}

\textbf{Three levels of redundancy.}
There exists three levels of redundancy in the diffusion process. The first type refers to the \textbf{step level} redundancy, which manifests as high similarity between consecutive timestep outputs. 
The second type is the \textbf{cfg level} redundancy, indicating that the outputs of conditional forward and unconditional forward within the same timestep are similar. 
The third type is \textbf{block level} redundancy, which means that the output of some transformer blocks at this timestep is similar to the output of the same block at the previous timestep. 
We adopt the relative L1 distance to characterize the similarity of two outputs, and the three-level redundancy is measured as follows:
\begin{equation}
\begin{aligned}
D^{step}_t &= \frac{||O^{step}_t - O^{step}_{t-1}||_1}{||O^{step}_{t-1}||_1} \\
D^{cfg}_t  &= \frac{||O^{uncond}_t - O^{cond}_{t}||_1}{||O^{cond}_{t}||_1} \\
D^{block_i}_t &= \frac{||O^{block_i}_t - O^{block_i}_{t-1}||_1}{||O^{block_i}_{t-1}||_1}, i \in [0,L)
\end{aligned}
\end{equation}
where $t$ denotes the sampling timestep index, advancing incrementally as the diffusion process progresses.
The smaller the $D$ value, the higher the similarity between two outputs. 
As presented in Figure~\ref{fig:diff}, we use different prompts to examine the different-level redundancy on Wan and HunyuanVideo.
Such redundancy offers opportunities for caching-based computation skipping to enhance inference efficiency.

\textbf{The dynamism of redundancy.}
In Figure~\ref{fig:diff}, the redundancy during the diffusion process shows strong dynamism.
Specifically, there are the following manifestations: 
(1) All of the three-level redundancy demonstrates strong correlations across timesteps: the initial distance value is relatively high, and gradually decreases and stabilizes. This indicates that the early diffusion stage is sensitive with low redundancy, thereby making it unsuitable for caching.
(2) The speed at which redundancy decreases varies among different prompts.
(3) The degree of redundancy varies with different levels. For example, in Wan, the cfg level redundancy always remains the strongest throughout the diffusion process. The redundancy of block 10 in the early diffusion stage is stronger than that of block 30, while it is weaker than block 30 in the later diffusion stage.
The above phenomenons necessitate an adaptive and unified hybrid caching mechanism.

\begin{figure}[t]
\centering
\includegraphics[width=1.0\columnwidth]{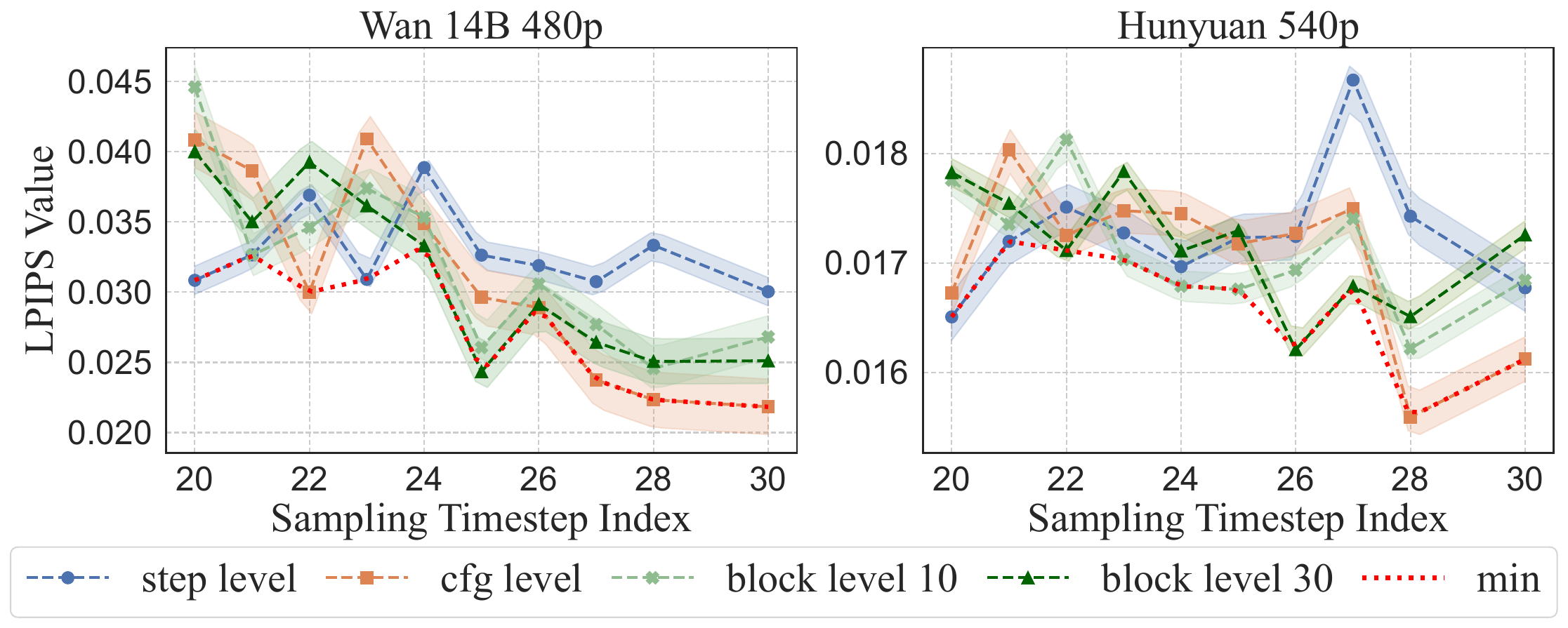}
\caption{Similarity metric compared with the original model using different cache strategies in different timesteps.}
\label{fig:three_level_cache_lpips}
\end{figure}

\begin{figure*}[t]  
\centering
\includegraphics[width=2\columnwidth]{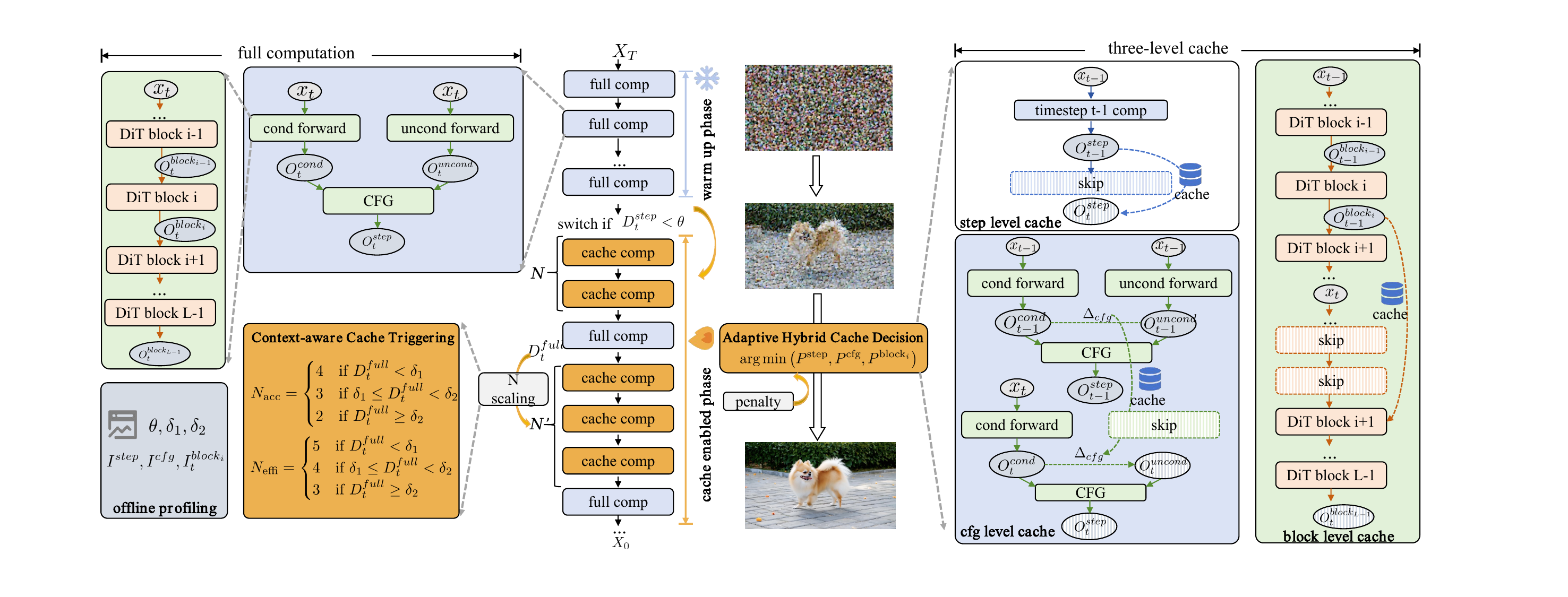}
\caption{The MixCache framework.}
\label{fig:method}
\end{figure*}

\textbf{Potential of three-level cache integration.}
We implement different granularity caching methods at each diffusion timestep, and adopt the LPIPS~\cite{LPIPS} metric to calculate the end-to-end similarity with the original video. The smaller the LPIPS value, the higher the similarity. 
As shown in Figure~\ref{fig:three_level_cache_lpips}, the optimal cache granularity varies dynamically across timesteps, suggesting that adaptive cache selection could enhance generation quality.

In order to better integrate different caching granularities, there are two core issues: (1) When will caching be triggered? That is, which timesteps enable caching and which perform full computation? (2) Which caching granularity should be selected for a given cache enabled timestep? 
To address the above issues, we propose the context-aware cache triggering strategy and adaptive hybrid cache decision strategy, respectively.
Integrating these two strategies, we propose the MixCache framework, aiming to achieve video DiT inference acceleration while maintaining a comparable quality of video generation. The overall MixCache framework is demonstrated in Figure~\ref{fig:method}.

\subsection{Context-aware Cache Triggering} \label{sec:when}
As the early diffusion stage is responsible for overall framework sketching, it is highly sensitive to interference at this time, as described in SRDiffusion~\cite{SRDiffusion}. This phenomenon can also be validated in Figure~\ref{fig:diff}, where in the initial diffusion process, the three-level redundancy is relatively low. 
Therefore, we perform full computation in the initial diffusion stage, and calculate $D^{step}_t$ between the output of the current timestep $t$ and that of the previous timestep $t-1$, and we call this stage as the \textit{warm up stage}. 
When $D^{step}_t$ is smaller than the predefined threshold $\theta$, established in the offline profiling process (detailed later), the warm up phase ends and enters the \textit{cache enabled phase}.

Once entering the cache enabled phase, in order to ensure the quality of video generation, it is necessary to perform full computation at certain intervals. 
A key issue is to determine which timesteps to perform full computation, and we refer to the number of cache enabled steps between two full computation timesteps as \textit{cache interval} (represented as $N$ in Figure~\ref{fig:method}). 
In Figure~\ref{fig:diff}, we observed that, after the warm up phase, the step level redundancy stabilizes at a certain value. 
Based on this, we propose an adaptive $N$ scaling strategy designed to dynamically monitor the deviation magnitude between the cached output and the ground-truth output, and automatically adjust the cache interval to maintain generation quality.
Specifically, after entering the cache enabled phase, we compare the output of two consecutive full computation, namely $D^{full}$, measured in relative L1 distance. 
When $D^{full}$ exceeds threshold $\delta_2$, this indicates the current cache granularity is too aggressive, requiring a reduction in the subsequent cache interval. Conversely, if it falls below $\delta_1$, the next cache interval should be increased.
In order to balance quality and efficiency, we provide two cache interval configurations, namely 2/3/4 ($N_{\text{acc}}$) and 3/4/5 ($N_{\text{effi}}$), which prioritize accuracy and efficiency respectively. 
The accuracy-prior cache interval $N_{\text{acc}}$ has a higher full computation frequency, resulting in better video generation quality:
\begin{equation}
\begin{aligned}
N_{\text{acc}} &=
  \begin{cases}
    4 & \text{if } D^{full}_{t} < \delta_1 \\
    3 & \text{if } \delta_1 \leq D^{full}_{t} < \delta_2 \\
    2 & \text{if } D^{full}_{t} \geq \delta_2
  \end{cases}
\end{aligned}
\end{equation}

\subsection{Adaptive Hybrid Cache Decision}  \label{sec:how}
\begin{figure*}[htbp]
    \centering
    \begin{subfigure}{0.37\textwidth}
        \centering
        \includegraphics[width=\linewidth]{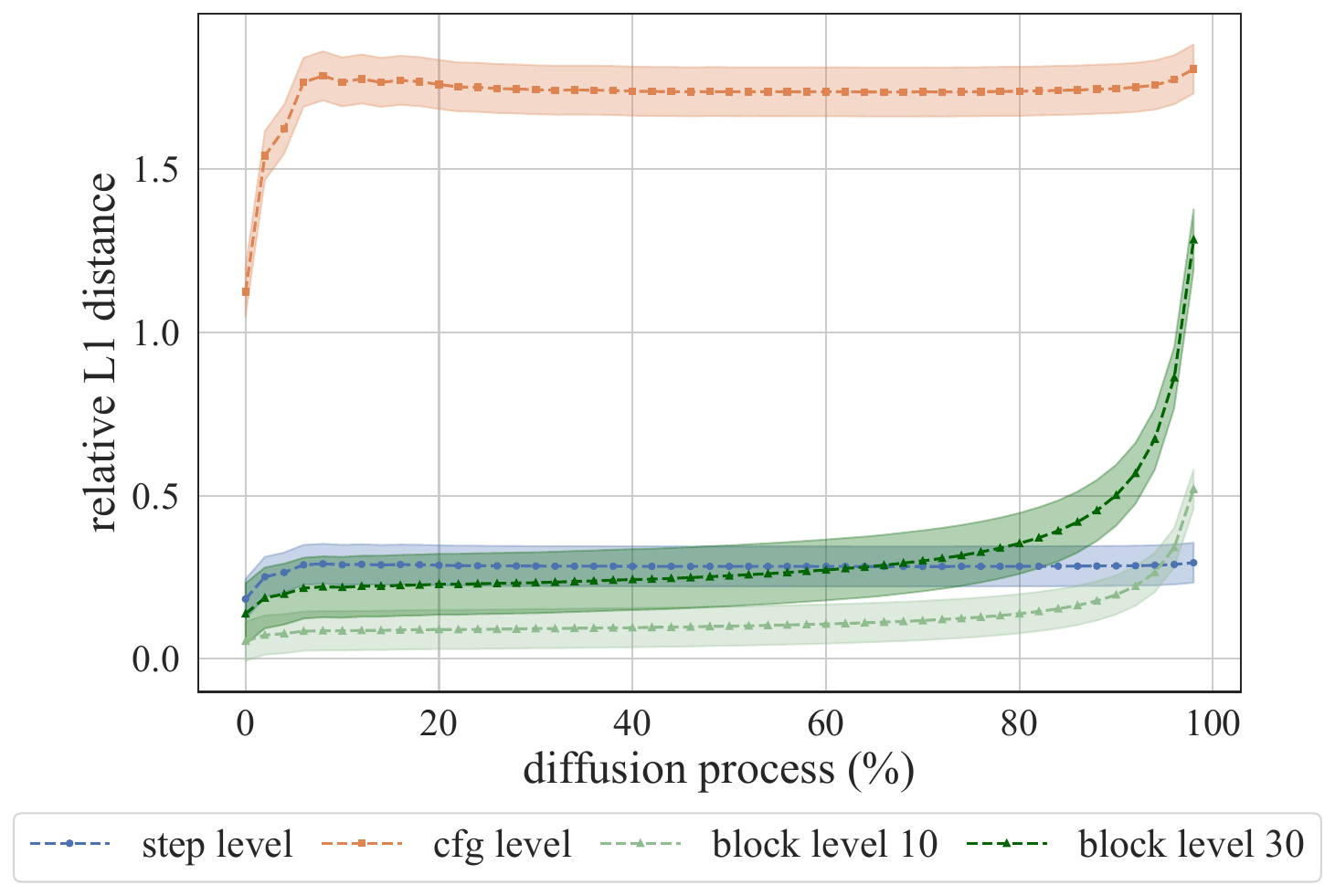}
    \end{subfigure}%
    \hfill
    \begin{subfigure}{0.62\textwidth}
        \centering
        \includegraphics[width=\linewidth]{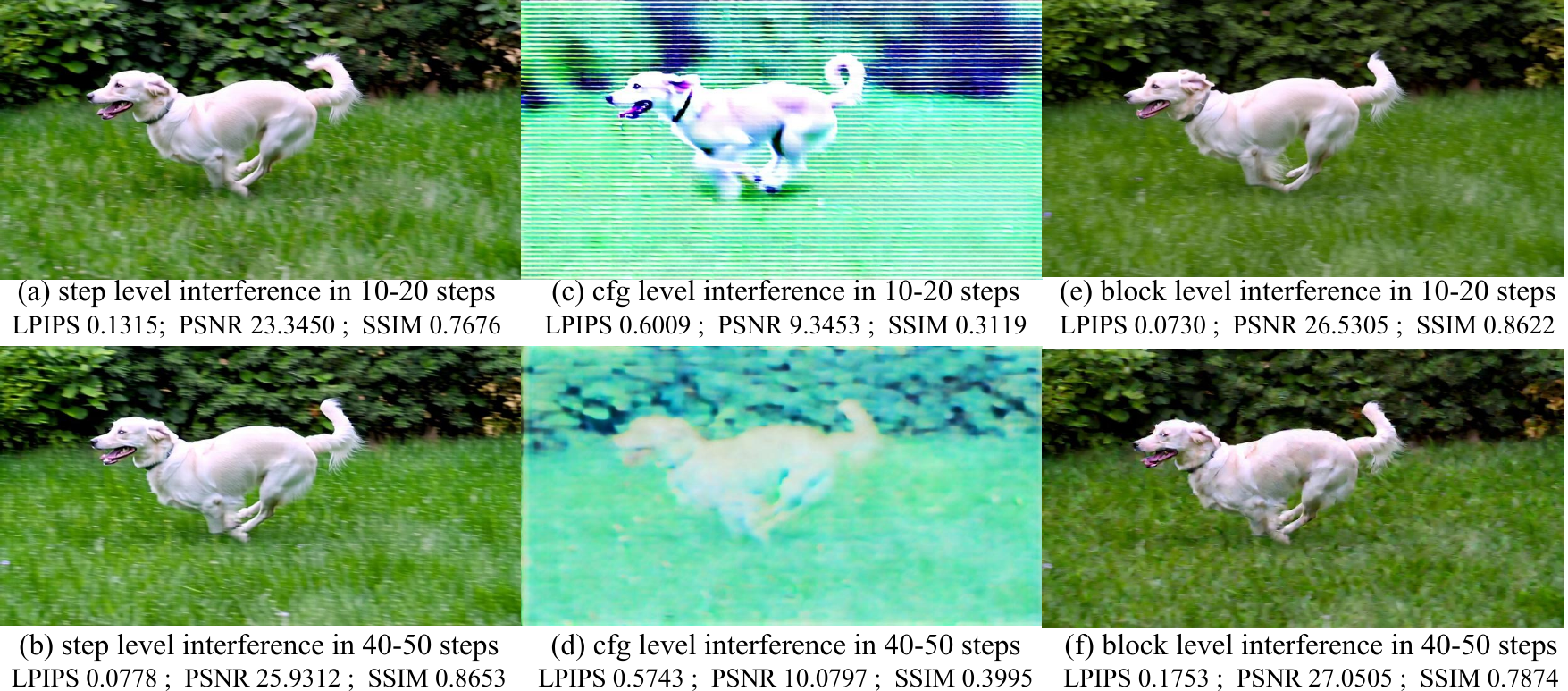}
    \end{subfigure}
    \caption{(a) left: Distance between the perturbed and original output, which performs as the impact indicator. (b) right: Visualization and quantitative metrics (LPIPS ↓ PSNR ↑ SSIM ↑) of different level interference at different diffusion stages.}
    \label{fig:impact}
\end{figure*}

\begin{figure}[t]
\centering
\includegraphics[width=1\columnwidth]{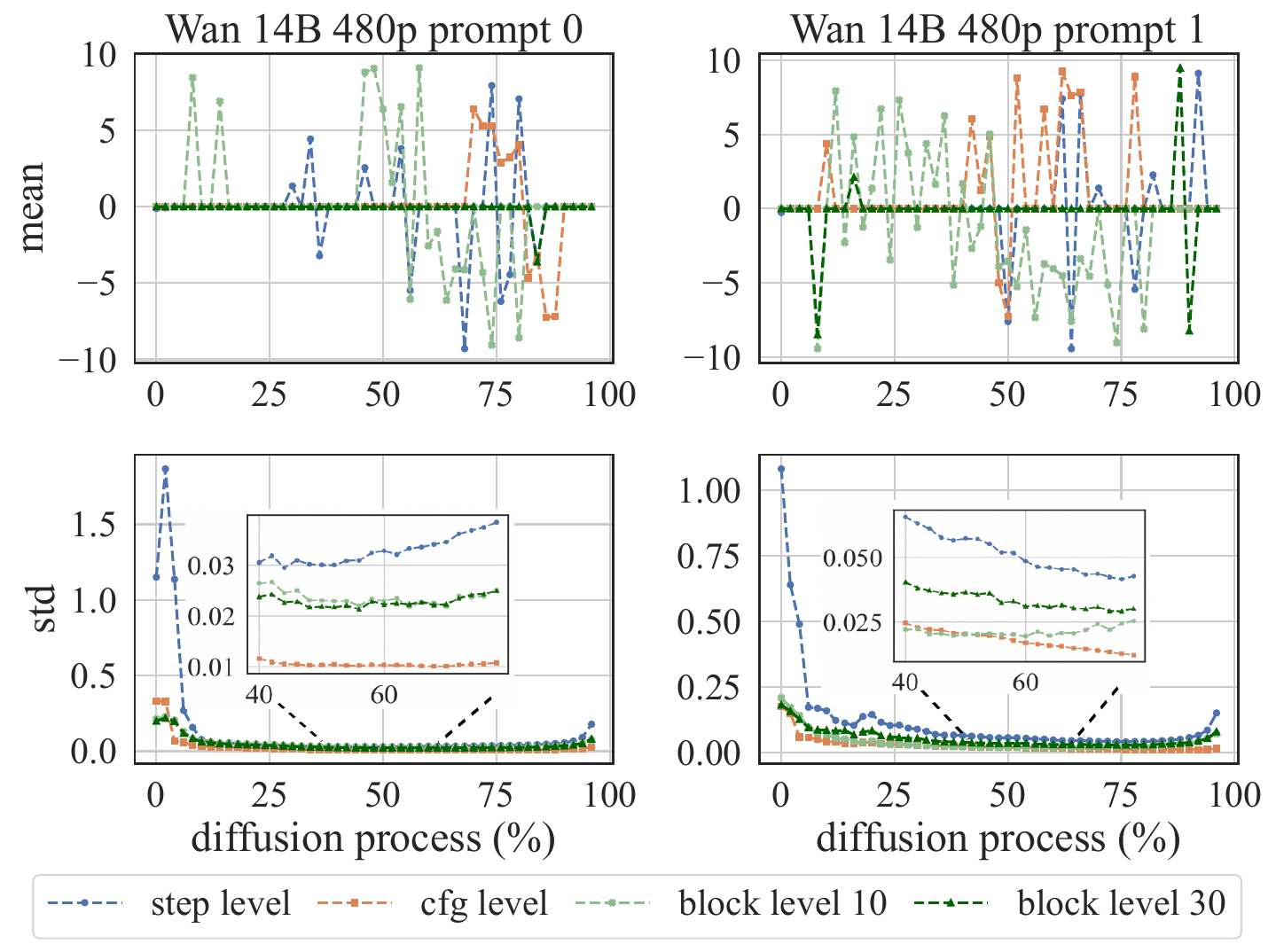}
\caption{The value of mean $\mu$ and standard deviation $\sigma$ of the distance tensor between the cached output and the original output.}
\label{fig:mean_and_std}
\vspace{-10pt}
\end{figure}

After identifying cache enabled timesteps, it is crucial to determine the specific caching granularity for a certain timestep within the three-level caching.
The limitation of prior works~\cite{TeaCache, FastCache, MagCache} is their exclusive focus on similarity of different cache methods while neglecting their differential impact on accuracy.
To assess the accuracy impact of different caching methods, we generate Gaussian distribution with predefined statistical parameters (mean $\hat{\mu}$, standard deviation $\hat{\sigma}$) and calculate the distance between perturbed and original outputs.
It should be noted that $\hat{\mu}$ and $\hat{\sigma}$ should align with the real parameters derived from the three-level caching.
Therefore, we firstly profile prompts and examine the $\mu$ and $\sigma$ of the distance tensor between the real output and the cached output.
As shown in Figure~\ref{fig:mean_and_std}, the $\mu$ value of the distance tensor is inconsistent across different prompts, and also varies at different timesteps. However, its values are fixed within a certain range. Therefore, we set the $\hat{\mu}$ value as the average of $\mu$ value across all timesteps.
In addition, for the $\hat{\sigma}$ value, except for the initial timesteps, it remains stable at a fix value. 
Therefore, we set the $\hat{\sigma}$ value as the average of the $\sigma$ value of the last 40 timesteps. 
We exclude the first 10 timesteps from analysis since these initial timesteps correspond to the warm up phase without applying caching. Consequently, their outputs lack statistical representativeness for evaluating the proposed methodology.

After determining the $\hat{\mu}$ and $\hat{\sigma}$ value, we examine the relative L1 distance between the perturbed output and the original output.
We examine for all timesteps and plot the results in Figure~\ref{fig:impact}(a). The smaller the distance, the smaller the impact of the interference on the actual results. This value can indicate the accuracy impact of performing a specific level cache at a specific timestep. 
From Figure~\ref{fig:impact}(a), it can be seen that the impact value of step level and cfg level remain relatively stable after the warm up phase.
However, the block level value exhibits time-dependent characteristics, where its initial value is low, and then increases in the later diffusion stage.
In addition, the compare across three levels reveals that, the cfg level interference exerts a substantially greater influence, with its value exceeding both step level and block level interference by an order of magnitude.
We present the visualized results and similarity metrics in Figure~\ref{fig:impact}(b). 
It can be seen that the video quality generated using cfg level interference is poor, indicating that this interference has a significant impact on the final results, which is consistent with the conclusion in Figure~\ref{fig:impact}(a). 
Therefore,  the values presented in Figure~\ref{fig:impact}(a) are employed as quantitative evaluation metrics for assessing accuracy impact, with the impact of step level and cfg level remaining constant ($I^{step}$, $I^{cfg}$), while the block level impact exhibit as a time-dependent and index-dependent function ($I^{block_i}_{t}$). \wyx{Thus, the offline Gaussian perturbation estimates the quantified impact value for each cache type using $\hat{\mu}$ and $\hat{\sigma}$ fit from measured cache-induced differences.} 

The similarity and accuracy impact jointly determine the final generated video quality. 
We prefer to employ the caching method with a lower similarity and a lower impact, as it indicates that such method exerts minimal influence on the generated video.
Therefore, we utilize the product of these two values to represent the final metric, measured as $P$ value.
After each cache enabled timestep, we calculate the similarity of the three levels and get $D^{step}_t$, $D^{cfg}_t$, $D^{block_i}_t$, and product them with the corresponding impact value.
Based on the greedy principle, we choose the cache method with the smallest $P$ as the cache method of the next cache enabled timestep. 
\begin{equation}
P_t^{\psi} = D^{\psi}_t \cdot I^{\psi}_t, \psi \in \{step, cfg, block_i\}
\end{equation}
\wyx{where i is one of the pre-defined block indices.}
\wyx{To prevent getting stuck in the same caching granularity within a cache interval, we introduce a \textit{penalty strategy}: if a caching granularity is used in one timestep, it is disabled in the next timestep. This proactive approach ensures flexibility and robustness, enabling systems to avoid local optima.}


\subsection{Unify it together: MixCache Framework}
The context-aware cache triggering strategy and adaptive hybrid cache decision strategy solve the issue of when to enable caching and which cache granularity to select, respectively. 
Unifying these two strategies, we obtain the MixCache framework. MixCache provides a training-free and adaptive hybrid cache mechanism that can effectively combine three three-level caching, achieving a balance between generation quality and inference speed. 


According to Figure~\ref{fig:diff}, the redundancy level of different models varies.
Given a specific model, MixCache first performs offline profiling to determine key hyper-parameters. We run 100 prompts through the diffusion process, recording $D^{step}$ at each timestep. Once $D^{step}$ stabilizes, which is identified when its value remains below a certain value compared to the previous five steps, we compute the average as $\theta$. The maximum and minimum deviations from $\theta$ after stabilization are set as $\delta_1$ and $\delta_2$. These values are averaged across all prompts and rounded as needed.
The profiling also determines $\hat{\mu}$ and $\hat{\sigma}$ for Gaussian interference, used to assess impact at different granularities. This process is performed once per model and enables efficient runtime deployment without extra overhead.

After the offline profiling process, it enters the runtime inference phase, and the execution flow is formalized in Algorithm~\ref{alg:adaptive_cache}. 
During runtime, one generation is partitioned into two distinct phases: the warm up phase and the cache enabled phase. 
In the warm up phase, the system exclusively executes full computations to maintain video quality, and this phase is ended by the context-aware cache triggering strategy to determine which timestep to enable caching (line 5). 
Once into the cache enabled phase,  MixCache uses the adaptive hybrid cache decision strategy to determine which of the three-level caching methods to use. 
There are two core mechanisms are activated: (1) the $N$ scaling strategy to regulate the frequency of cache interval (line 11), and (2) the adaptive hybrid cache decision strategy (line 23-24) that dynamically determines the optimal caching granularity for each timestep. 

\begin{algorithm}[t]
\caption{The execution flow of the MixCache framework}
\label{alg:adaptive_cache}
\begin{algorithmic}[1]
\State \textbf{Offline Profiling}: Obtain $\theta$, $\delta_1$, $\delta_2$, $I^{\text{step}}$, $I^{\text{cfg}}$,  $I^{\text{block}_i}_t$
\State \ \textbf{Initialize}: cache interval $N$, cnt = 0
\For{each sampling timestep $t \in \{0,1,2,\dots,T-1\}$}
    \State Compute $D^{\text{step}}_t$, $D^{\text{cfg}}_t$, $D^{\text{block}_i}_t$
    \If{$D^{\text{step}}_t \geq \theta$} \Comment{Warm-up phase}
        \State Perform full computation
    \Else \Comment{Cache enabled phase}
        \State $\text{cnt} \gets (\text{cnt} + 1) $\%$ \text{N}$
        \If{$\text{cnt} == 0$}
            \State Perform full computation
            \State Compute $D^{\text{full}}$ and scale $N$
        \ElsIf{cache mode is step level}
            \State O$_t$ $\gets$ O$_{t-1}$
        \ElsIf{cache mode is cfg level}
            \State O$^{cond}_t$ $\gets$ conditional forward output
            \State O$^{uncond}_t$ $\gets$ O$^{cond}_t$ + $\Delta_{cfg}$
            \State O$_t$ $\gets$ CFG(O$^{cond}_t$, O$^{uncond}_t$)
        \ElsIf{cache mode is block level $i$}
            \State Input for block $i+1$ $\gets$ O$_{t-1}^{\text{block}_i}$
            \State Execute DiT block[i+1:]
            \State O$^{uncond}_t$ $\gets$ O$^{cond}_t$ + $\Delta_{cfg}$
            \State O$_t$ $\gets$ CFG(O$^{cond}_t$, O$^{uncond}_t$)
        \EndIf
        \State $P_t^{\psi} \gets D^{\psi}_t \cdot I^{\psi}, \psi \in \{step, cfg, block\}$
        \State cache mode $\gets$ $arg\min\left(P^{\text{step}}, P^{\text{cfg}}, P^{\text{block}_i}\right)$
    \EndIf
\EndFor
\end{algorithmic}
\end{algorithm}

\vspace{25pt}
\section{Experiments}

\begin{table}[t]
\resizebox{0.49\textwidth}{!}{%
\centering
\setlength{\tabcolsep}{2.5pt}
\begin{tabular}{ccccccc}
\hline
\multicolumn{1}{c|}{}                                         & \multicolumn{4}{c|}{\textbf{Visual Quality}}                                                                                                                                                               & \multicolumn{2}{c}{\textbf{Efficiency}}                            \\ \cline{2-7} 
\multicolumn{1}{c|}{\multirow{-2}{*}{\textbf{Methods}}}       & \multicolumn{1}{c}{\textbf{Vbench↑}}      & \multicolumn{1}{c}{\textbf{LPIPS↓}}         & \multicolumn{1}{c}{\textbf{PSNR↑}}           & \multicolumn{1}{c|}{\textbf{SSIM↑}}          & \multicolumn{1}{c}{\textbf{Latency}}              & \textbf{Speedup}         \\ \hline
\multicolumn{7}{c}{\textbf{Wan 14B} (832×480, 5s, 81 frames, T = 50)}                                                                                                                                                                                                                                             \\ \hline
\rowcolor{gray!15} \multicolumn{1}{c|}{\textbf{original}}                        & \multicolumn{1}{c}{84.05}                        & \multicolumn{1}{c}{-}                        & \multicolumn{1}{c}{-}                        & \multicolumn{1}{c|}{-}                        & \multicolumn{1}{c}{900 s}      & -             \\ 
\multicolumn{1}{c|}{Teacache$_{0.1}$}              & \multicolumn{1}{c}{\textbf{84.01}}               & \multicolumn{1}{c}{0.147}                    & \multicolumn{1}{c}{22.17}                    & \multicolumn{1}{c|}{0.786}                    & \multicolumn{1}{c}{849 s}                         & 1.06 $\times$                   \\ 
\multicolumn{1}{c|}{Teacache$_{0.14}$}             & \multicolumn{1}{c}{83.95}                        & \multicolumn{1}{c}{0.244}                    & \multicolumn{1}{c}{18.60}                    & \multicolumn{1}{c|}{0.688}                    & \multicolumn{1}{c}{612 s}                         & 1.47 $\times$                     \\ 
\multicolumn{1}{c|}{FasterCache}                     & \multicolumn{1}{c}{83.40}                        & \multicolumn{1}{c}{0.140}                    & \multicolumn{1}{c}{23.26}                    & \multicolumn{1}{c|}{0.796}                    & \multicolumn{1}{c}{633 s}                         & 1.42 $\times$                      \\ 
\multicolumn{1}{c|}{BlockDance$_4$}            & \multicolumn{1}{c}{83.48}                        & \multicolumn{1}{c}{0.129}                    & \multicolumn{1}{c}{\textbf{24.01}}                    & \multicolumn{1}{c|}{0.811}                    & \multicolumn{1}{c}{679 s}                         & 1.29 $\times$                     \\ 
\multicolumn{1}{c|}{PAB$^8_{100,800}$}                   & \multicolumn{1}{c}{83.00}                        & \multicolumn{1}{c}{0.166}                    & \multicolumn{1}{c}{22.29}                    & \multicolumn{1}{c|}{0.772}                    & \multicolumn{1}{c}{717 s}                         & 1.25 $\times$                     \\ 
\rowcolor{gray!15} \multicolumn{1}{c|}{\textbf{MixCache$_{\text{acc}}$}}          & \multicolumn{1}{c}{83.97}                        & \multicolumn{1}{c}{\textbf{0.124}}           & \multicolumn{1}{c}{23.45}                    & \multicolumn{1}{c|}{\textbf{0.814}}           & \multicolumn{1}{c}{528 s}                         & 1.70 $\times$                     \\ 
\rowcolor{gray!15} \multicolumn{1}{c|}{\textbf{MixCache$_{\text{effi}}$}}         & \multicolumn{1}{c}{83.90}                        & \multicolumn{1}{c}{0.132}                    & \multicolumn{1}{c}{22.94}                    & \multicolumn{1}{c|}{0.804}                    & \multicolumn{1}{c}{\textbf{465 s}}                & \textbf{1.94 $\times$}          \\ \hline
\multicolumn{7}{c}{\textbf{Wan 14B} (1280*720, 5s, 81 frames, T = 50)}                                                                                                                                                                                        \\ \hline
\rowcolor{gray!15} \multicolumn{1}{c|}{original}                        & \multicolumn{1}{c}{83.66}                        & \multicolumn{1}{c}{-}                        & \multicolumn{1}{c}{-}                          & \multicolumn{1}{c|}{-}                       & \multicolumn{1}{c}{3168 s}         & {-} \\ 
\multicolumn{1}{c|}{Teacache$_{0.1}$}              & \multicolumn{1}{c}{83.75}               & \multicolumn{1}{c}{0.148}                    & \multicolumn{1}{c}{22.58}                    & \multicolumn{1}{c|}{0.813}                    & \multicolumn{1}{c}{2988 s}                        & 1.06 $\times$                   \\ 
\multicolumn{1}{c|}{Teacache$_{0.14}$}             & \multicolumn{1}{c}{\textbf{83.78}}                        & \multicolumn{1}{c}{0.237}                    & \multicolumn{1}{c}{19.24}                    & \multicolumn{1}{c|}{0.732}                    & \multicolumn{1}{c}{2155 s}                        & 1.47 $\times$                   \\ 
\multicolumn{1}{c|}{FasterCache}                     & \multicolumn{1}{c}{83.12}                        & \multicolumn{1}{c}{0.156}                    & \multicolumn{1}{c}{22.97}                    & \multicolumn{1}{c|}{0.806}                    & \multicolumn{1}{c}{2230 s}                        & 1.42 $\times$                    \\ 
\multicolumn{1}{c|}{BlockDance$_4$}            & \multicolumn{1}{c}{83.43}                        & \multicolumn{1}{c}{0.136}                    & \multicolumn{1}{c}{23.70}                    & \multicolumn{1}{c|}{0.820}                    & \multicolumn{1}{c}{2455 s}                        & 1.29 $\times$                   \\ 
\multicolumn{1}{c|}{PAB$^8_{100,800}$}             & \multicolumn{1}{c}{83.32}                             & \multicolumn{1}{c}{0.195}                         & \multicolumn{1}{c}{21.41}                         & \multicolumn{1}{c|}{0.775}                         & \multicolumn{1}{c}{2534 s}                        & 1.25 $\times$                   \\
\rowcolor{gray!15} \multicolumn{1}{c|}{\textbf{MixCache$_{\text{acc}}$}}          & \multicolumn{1}{c}{83.74}                        & \multicolumn{1}{c}{\textbf{0.132}}           & \multicolumn{1}{c}{\textbf{23.88}}           & \multicolumn{1}{c|}{\textbf{0.824}}           & \multicolumn{1}{c}{1951 s}                        & 1.62 $\times$                    \\ 
\rowcolor{gray!15} \multicolumn{1}{c|}{\textbf{MixCache$_{\text{effi}}$}}         & \multicolumn{1}{c}{83.70}                        & \multicolumn{1}{c}{0.146}                    & \multicolumn{1}{c}{22.81}                    & \multicolumn{1}{c|}{0.815}                    & \multicolumn{1}{c}{\textbf{1742 s}}               & \textbf{1.82 $\times$}           \\ \hline
\multicolumn{7}{c}{\textbf{HunyuanVideo} (960×544, 5s, 129 frames, T = 50)}                                                                                                                                                                                                                                                                                \\ \hline
\rowcolor{gray!15} \multicolumn{1}{c|}{original}                        & \multicolumn{1}{c}{81.13}                        & \multicolumn{1}{c}{-}                        & \multicolumn{1}{c}{-}                        & \multicolumn{1}{c|}{-}                        & \multicolumn{1}{c}{2289 s}             &       -        \\ 
\multicolumn{1}{c|}{Teacache$_{0.1}$}              & \multicolumn{1}{c}{80.87}                        & \multicolumn{1}{c}{0.247}                    & \multicolumn{1}{c}{17.57}                    & \multicolumn{1}{c|}{0.734}                    & \multicolumn{1}{c}{1421 s}                              &          1.61 $\times$                \\ 
\multicolumn{1}{c|}{BlockDance$_4$}            & \multicolumn{1}{c}{80.93}                             & \multicolumn{1}{c}{0.051}                         & \multicolumn{1}{c}{\textbf{28.80}}                         & \multicolumn{1}{c|}{0.897}                         & \multicolumn{1}{c}{1646 s}                              &                   1.39 $\times$       \\ 
\multicolumn{1}{c|}{PAB$^8_{100,800}$}                             & \multicolumn{1}{c}{80.64}                             & \multicolumn{1}{c}{0.066}                         & \multicolumn{1}{c}{28.56}                         & \multicolumn{1}{c|}{0.901}                         & \multicolumn{1}{c}{1847 s}                              &         1.24 $\times$                 \\ 
\rowcolor{gray!15} \multicolumn{1}{c|}{\textbf{MixCache$_{\text{acc}}$}}          & \multicolumn{1}{c}{\textbf{81.05}}                             & \multicolumn{1}{c}{\textbf{0.047}}                         & \multicolumn{1}{c}{28.37}                         & \multicolumn{1}{c|}{\textbf{0.921}}                         & \multicolumn{1}{c}{1240 s}                              &            1.84 $\times$              \\ 
\rowcolor{gray!15} \multicolumn{1}{c|}{\textbf{MixCache$_{\text{effi}}$}}         & \multicolumn{1}{c}{80.98}                             & \multicolumn{1}{c}{0.060}                         & \multicolumn{1}{c}{26.86}                         & \multicolumn{1}{c|}{0.906}                         & \multicolumn{1}{c}{\textbf{1151 s}}                              &               \textbf{1.97} $\times$           \\ \hline
\multicolumn{7}{c}{\textbf{CogVideoX 5B} (820×480, 6s, 49 frames, T = 50)}                                                                                                                                                                                                                                                                            \\ \hline
\rowcolor{gray!15} \multicolumn{1}{c|}{\textbf{original}}                        & \multicolumn{1}{c}{80.89}                        & \multicolumn{1}{c}{-}                        & \multicolumn{1}{c}{-}                        & \multicolumn{1}{c|}{-}                        & \multicolumn{1}{c}{443 s}           & -              \\ 
\multicolumn{1}{c|}{Teacache$_{0.1}$}              & \multicolumn{1}{c}{\textbf{80.15}}                        & \multicolumn{1}{c}{0.239}                    & \multicolumn{1}{c}{20.42}                    & \multicolumn{1}{c|}{0.741}                    & \multicolumn{1}{c}{289 s}                         & 1.53 $\times$                   \\ 
\multicolumn{1}{c|}{BlockDance$_4$}            & \multicolumn{1}{c}{74.34}                        & \multicolumn{1}{c}{0.349}                    & \multicolumn{1}{c}{24.76}                    & \multicolumn{1}{c|}{0.750}                    & \multicolumn{1}{c}{348 s}                         & 1.27 $\times$                   \\ 
\multicolumn{1}{c|}{PAB$^8_{100,800}$}                   & \multicolumn{1}{c}{78.67}                        & \multicolumn{1}{c}{0.235}                    & \multicolumn{1}{c}{21.69}                    & \multicolumn{1}{c|}{0.770}                    & \multicolumn{1}{c}{293 s}                         & 1.51 $\times$                   \\ 
\rowcolor{gray!15} \multicolumn{1}{c|}{\textbf{MixCache$_{\text{acc}}$}}          & \multicolumn{1}{c}{80.10}                             & \multicolumn{1}{c}{\textbf{0.089}}                         & \multicolumn{1}{c}{\textbf{34.33}}                         & \multicolumn{1}{c|}{\textbf{0.892}}                         & \multicolumn{1}{c}{287 s}                              &         1.54 $\times$                 \\ 
\rowcolor{gray!15} \multicolumn{1}{c|}{\textbf{MixCache$_{\text{effi}}$}}         & \multicolumn{1}{c}{\textbf{80.15}}                             & \multicolumn{1}{c}{0.160}                         & \multicolumn{1}{c}{26.86}                         & \multicolumn{1}{c|}{0.880}                         & \multicolumn{1}{c}{\textbf{256 s}}                              &           \textbf{1.73} $\times$               \\ \hline
\end{tabular}
}
\caption{Comparison of efficiency and visual quality.}
\label{tab:res}
\end{table}

\begin{figure*}[h]
\centering
\includegraphics[width=2\columnwidth]{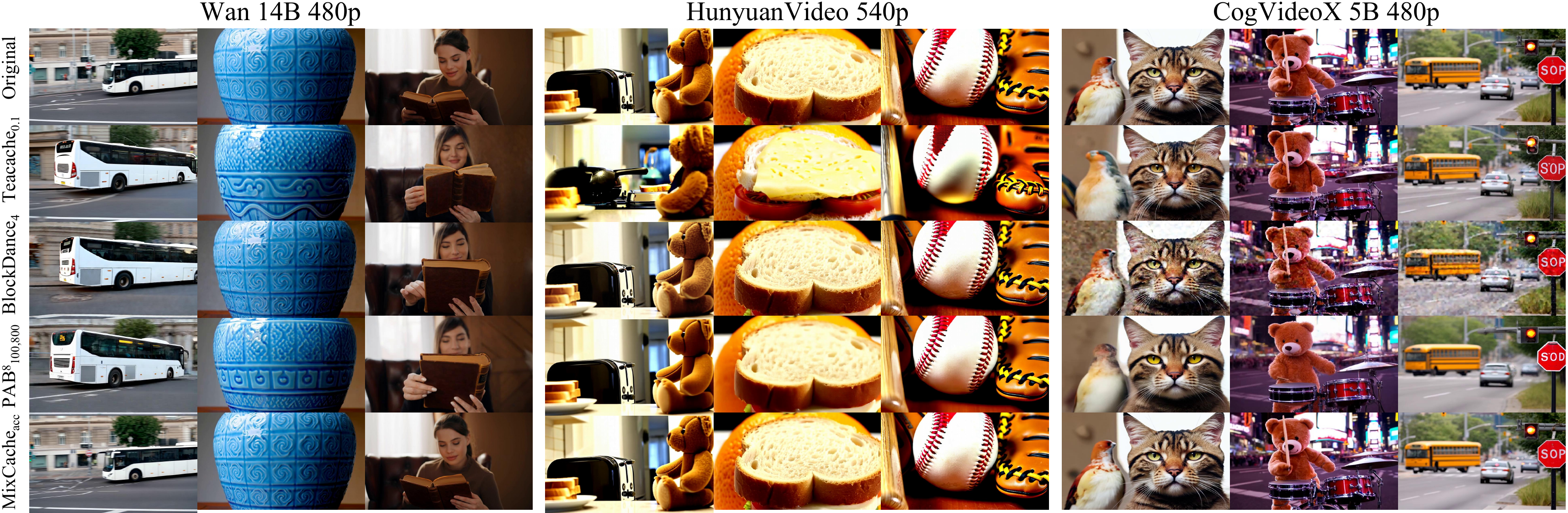}
\caption{Visual quality comparison. MixCache delivers high quality and maintains consistency with original results.} 
\label{fig:vis_baseline}
\end{figure*}

\subsection{Settings}
\textbf{Base Models and Baselines.}
We evaluate MixCache on Wan2.1 14B~\cite{Wan}, HunyuanVideo~\cite{Hunyuanvideo} and CogVideoX 5B~\cite{CogvideoX}. 
For baseline methods, we choose Teacache~\cite{TeaCache}, FasterCache~\cite{FasterCache}, BlockDance~\cite{BlockDance} and PAB~\cite{PAB}, all of which are specifically designed to accelerate DiT models through caching.
Among them, Teacache adopts step level cache, FasterCache combines cfg level and block level cache, and BlockDance and PAB employ block level cache. 
We adjust the parameters of each baseline to achieve a balance between speed and quality. 
Specifically, Teacache$_{0.1}$ indicates that the value of \texttt{L1\_distance\_thresh} from the open-source implementation is set to 0.1, and 
the implementation of PAB is adopted on HuggingFace Diffusers, where PAB$^8_{100,800}$ denotes a configuration of \texttt{block\_skip\_range}=8 and \texttt{timestep\_skip\_range}=$[100,800]$.
For BlockDance$_N$, following the official setting, we set the cache block index as 20 and set the first 40\% of denoising steps as warm up steps that disable caching, and evenly divide the remaining 60\% of denoising steps into several groups, each comprising $N$ steps.

\textbf{Evaluation Metrics.}
To evaluate the quality of video generation, we employ VBench~\cite{Vbench}, a widely-adopted comprehensive benchmarking suite for evaluating video generation.
Based on VBench standard prompt set, we use Qwen2.5-14B-Instruct~\cite{qwen2} to extend all prompts to enhance the video quality, and generate 5 videos with different seeds for each prompt. 
In addition, we report LPIPS~\cite{LPIPS}, SSIM~\cite{SSIM} and PSNR for quality comparison. 
For efficiency evaluation, we quantify the average inference latency per prompt as the primary performance metric.

\textbf{Experiment Settings.} 
For the main results, we generated 4720 videos for each set of results based on VBench. These were processed in data parallel across a public cloud instance with 64 NVIDIA A800 (80GB) GPUs, and the whole experiment takes nearly a month to complete.
Except the main results that use all prompts of Vbench, we randomly sample 200 prompts from VBench to conduct ablation experiments, each using 1 seed for generation. 

\textbf{Implementation Details.}
\wyx{All experiments conduct full attention inference.
We pre-defined block index as (10,20,30) as block caching candidates, which performs well for most models.
The profiled hyper-parameters $\theta$ and ($\delta_1$, $\delta_2$) for Wan and HunyuanVideo are 0.1, (0.05, 0.1), respectively, and those for CogVideoX are 0.1, (0.3, 0.4) respectively.}

\subsection{Main Results}
\noindent \textbf{Quantitative Comparison.} 
Table~\ref{tab:res} presents a quantitative comparison of MixCache with baselines in terms of efficiency and visual quality. 
The results highlight that, MixCache consistently demonstrates robust acceleration efficiency and maintains visually compelling quality across diverse base models, baselines, and resolutions. 

\noindent \textbf{Visual Comparison.} 
Figure~\ref{fig:vis_baseline} compares the videos generated by MixCache against those by the original model and baselines. 
The results demonstrate that MixCache can effectively preserve the original semantics and fine details. 

\vspace{15pt}
\subsection{Ablation Study}
\begin{figure}[t]
\centering
\includegraphics[width=1.0\columnwidth]{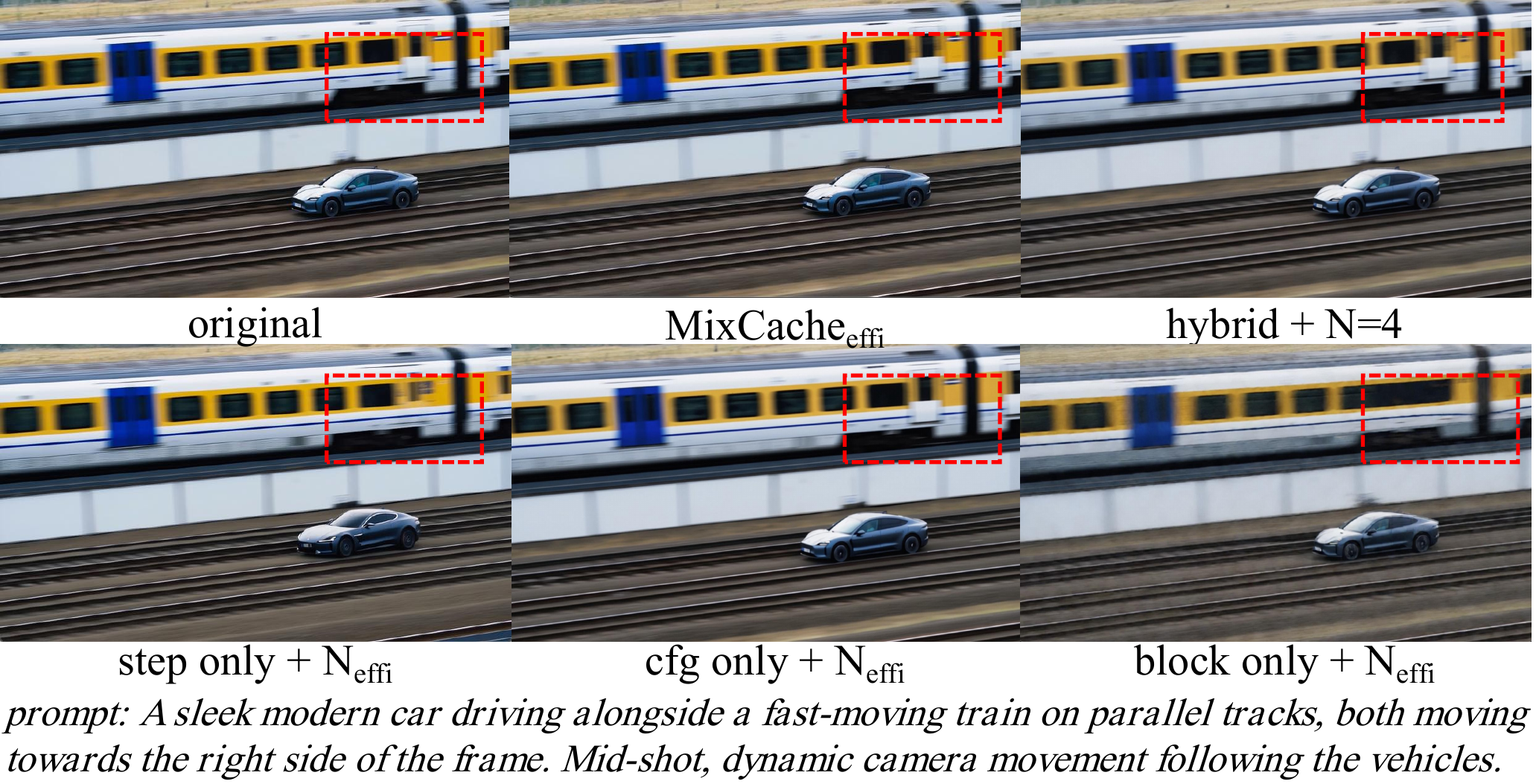}
\caption{Visualization of ablation study on Wan 14B 480p.}
\label{fig:vis_aba}
\end{figure}

In the ablation study, we investigate the effects of $N$ scaling strategy, penalty strategy and hybrid three-level caching respectively in terms of the quality and efficiency. 
The quantitative results are shown in Table~\ref{tab:aba} and the corresponding visualization results are presented in Figure~\ref{fig:vis_aba}. 
Among them, ``hybrid'' refers to using the adaptive hybrid cache decision strategy to determine the cache method for each cache enabled step, while ``step/cfg/block only'' refers to using only a specific caching method in all cache enabled steps. ``N=4'' represents a fixed cache interval of 4, and ``$N_{\text{effi}}$'' represents enabling the $N$ scaling strategy that prioritizes efficiency. 
It can be seen that MixCache outperforms all ablation experiments in both quality and efficiency metrics, indicating that combining dynamic $N$ scaling, penalty strategy and  hybrid three-level caching can achieve better generation quality and inference efficiency. 

\begin{table}[ht]
\resizebox{0.48\textwidth}{!}{%
\setlength{\tabcolsep}{1mm}
\begin{tabular}{c|ccc|cc}
\hline
                 \textbf{Method}             & \textbf{LPIPS ↓} & \textbf{PSNR ↑}  & \textbf{SSIM ↑} & \textbf{Latency} & \textbf{Speedup} \\ \hline
\rowcolor{gray!15} \textbf{original}               & -                & -                & -               & 900 s            & -                \\ 
\textbf{hybrid + N=4}           & 0.082           & 25.74          & 0.854          & 593 s            & 1.51 $\times$           \\ 
\textbf{step only + $N_{\text{effi}}$} & 0.147         & 22.25          & 0.775          & 545 s            & 1.65 $\times$           \\
\textbf{cfg only+ $N_{\text{effi}}$}   & 0.080           & 25.90         & 0.853          & 623 s            & 1.44 $\times$            \\ 
\textbf{block only+ $N_{\text{effi}}$} & 0.114          & 24.62         & 0.809          & 537 s            & 1.67 $\times$           \\ 
\textbf{MixCache$_{\text{effi}}$ wo. pen} & 0.102  & 23.75 & 0.796 & 525 s   & 1.71 x  \\ 
\rowcolor{gray!15} \textbf{MixCache$_{\text{effi}}$ wi. pen} & \textbf{0.079}  & \textbf{25.91} & \textbf{0.858} & \textbf{465 s}   & \textbf{1.94 x}  \\ 
\hline
\end{tabular}
}
\caption{The ablation study of $N$ scaling strategy, penalty strategy and three-level caching on Wan 14B 480p.}
\label{tab:aba}
\end{table}

\subsection{Adaptability}
\begin{figure}[t]
\centering
\includegraphics[width=1.0\columnwidth]{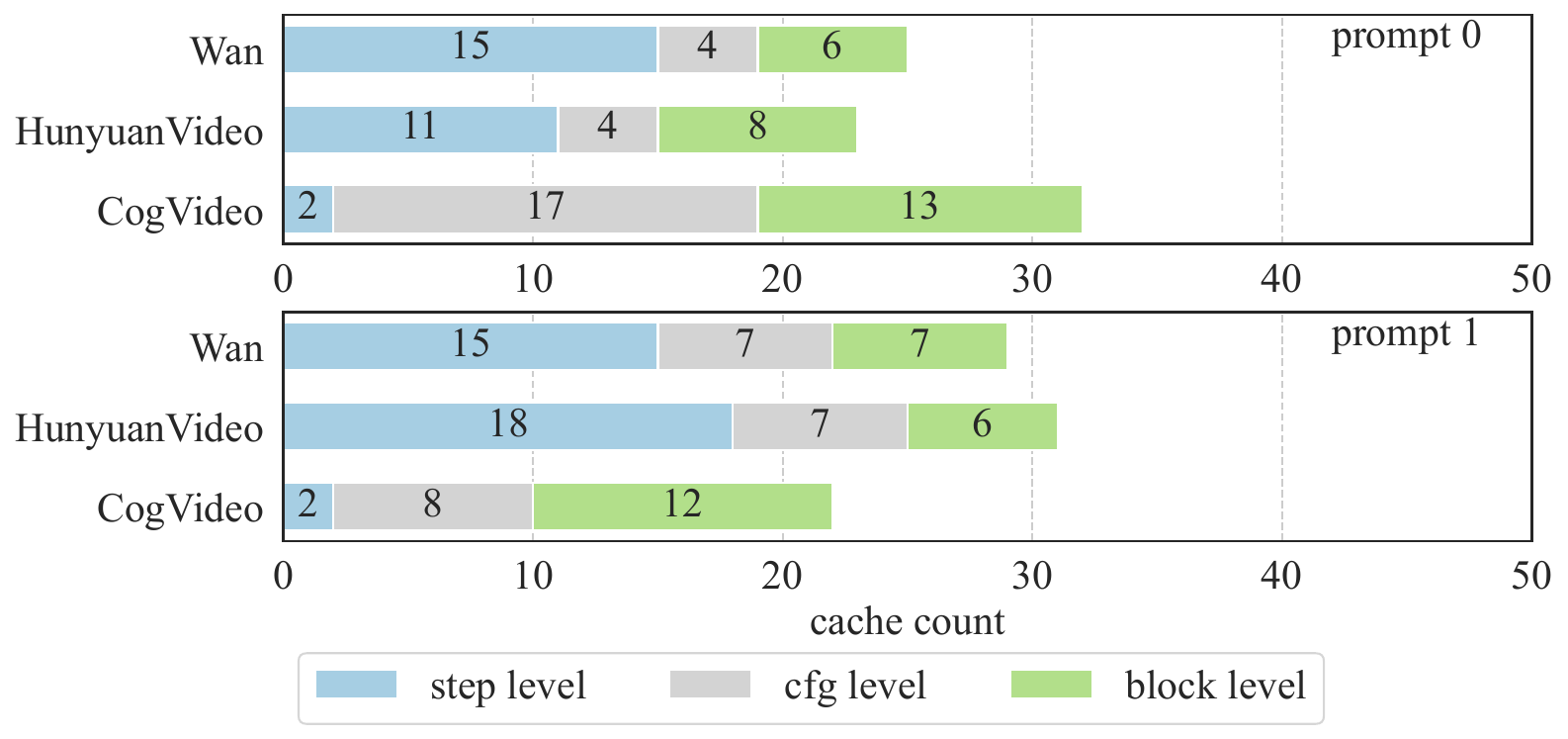}
\caption{Distribution of three-level caching on two prompts.}
\label{fig:propotion}
\end{figure}

As shown in Figure~\ref{fig:propotion}, we present the distribution of three-level caching across three models, evaluated on two distinct prompts. Through intra-model comparison across different prompts, it is evident that both the number of cache enabled timesteps and the distribution across three levels exhibit significant variations. 
Inter-model analysis reveals distinct cache utilization patterns. 
For instance, CogVideoX exhibits relatively fewer step level caching compared to other models,which attributes to its unique architectural property. 
These observations highlight the inherent adaptability of MixCache in accommodating diverse model architectures and prompts through context-aware resource allocation.

\begin{figure}[t]
\centering
\includegraphics[width=1.0\columnwidth]{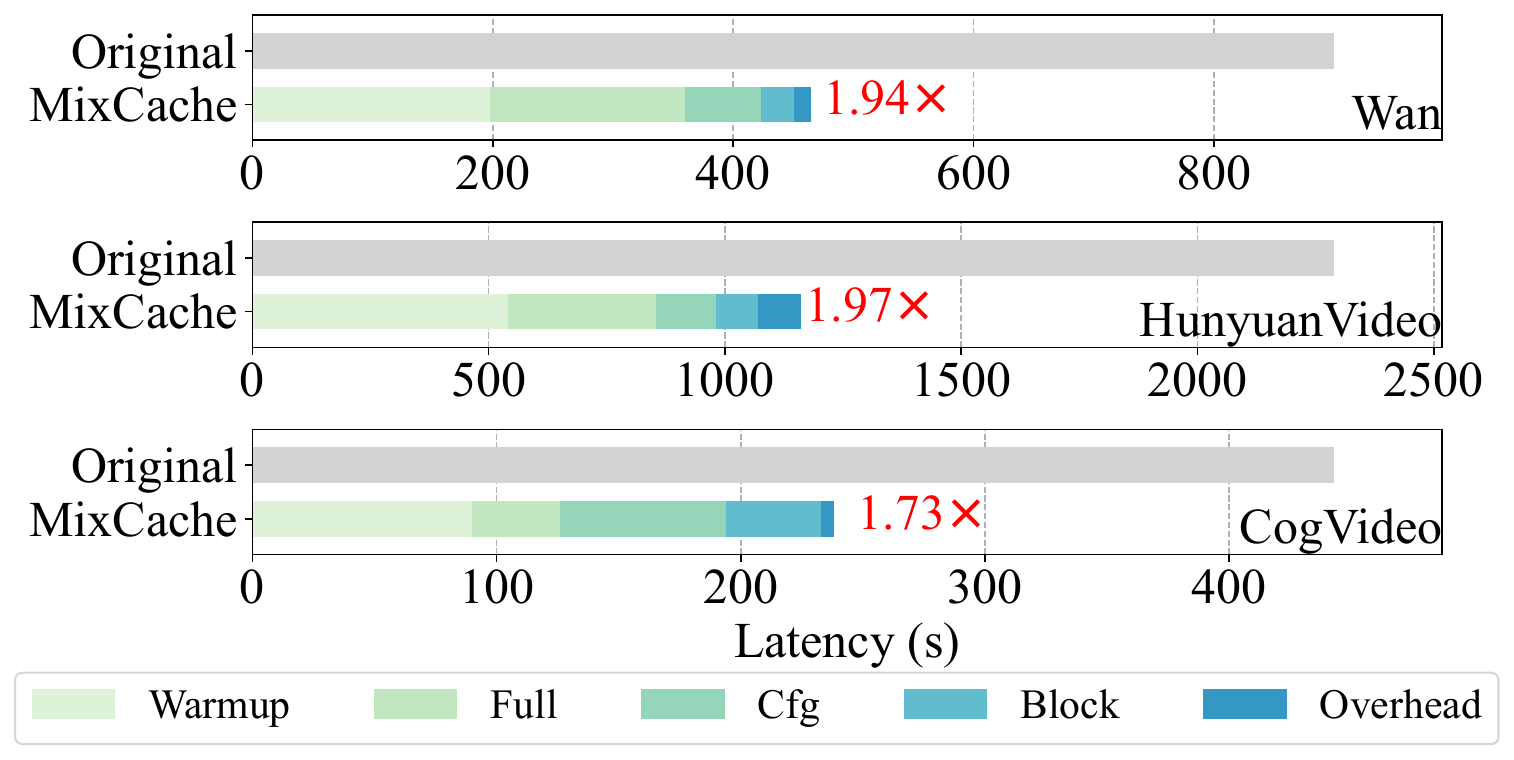}
\caption{Latency breakdown on three models.}
\label{fig:breakdown}
\end{figure}

\wyx{\subsection{Latency Breakdown}}
Figure~\ref{fig:breakdown} shows the latency breakdown of MixCache$_{\text{effi}}$ on three models. 
Among them, ``Full'' indicates the time for performing full computations at adaptive intervals after entering the cache enabled phase.
``Overhead'' is primarily used for determining cache intervals, calculating $D$ values, and selecting cache modes for each step. 
As step-level caching skips all the computation of that step, it is not included in the breakdown. 
As can be seen, MixCache's performance benefit primarily comes from significantly reducing full computation time with minimal overhead.

\begin{figure}[h]
\centering
\includegraphics[width=1.0\columnwidth]{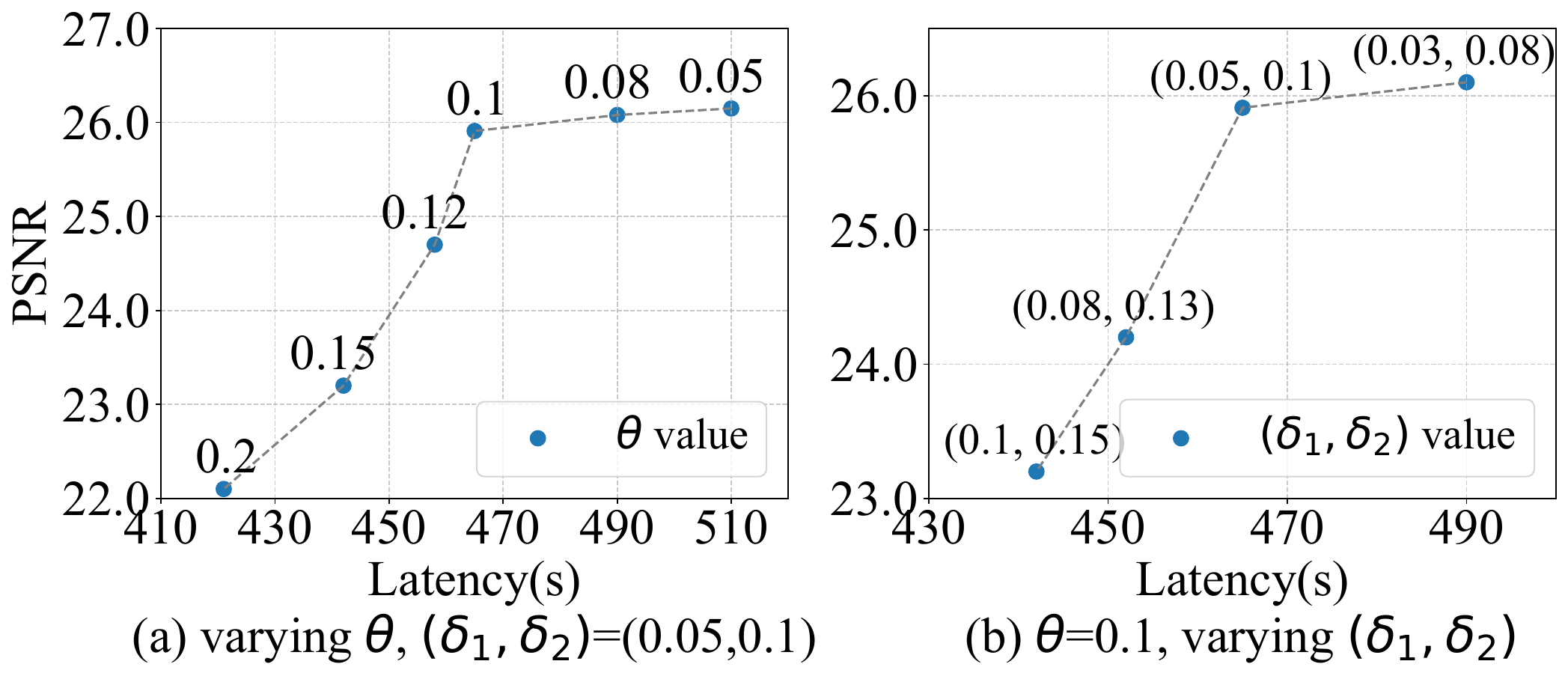}
\caption{Quality-latency tradeoff under different hyper-parameter configurations on Wan 14B 480p.}
\label{fig:hyper}
\vspace{-15pt}
\end{figure}

\wyx{\subsection{Hyperparameter Selection}}
We conduct experiments under different hyperparameter configurations on a wan 14B 480p with 200 prompts. 
Figure~\ref{fig:hyper} plots the generated quality (indicated by PSNR ↑) and latency for: (a) varying $\theta$ with ($\delta_1$, $\delta_2$)=(0.05,0.1) and (b) varying ($\delta_1$, $\delta_2$) with $\theta$=0.1. 
As can be seen, under different configurations, there is a trade-off between quality and latency. 
Specifically, Figure~\ref{fig:hyper}(a) shows that, when fixed ($\delta_1$, $\delta_2$)=(0.05,0.1), while $\theta$=0.08 or 0.05 has a slight improvement in generated quality compared to $\theta$=0.1, it suffers from a significant disadvantage in latency. 
This phenomenon can also be seen in Figure~\ref{fig:hyper}(b). 
Therefore, considering the trade-off between quality and speed, we choose the configuration with $\theta$=0.1 and ($\delta_1$, $\delta_2$)=(0.05,0.1).

\wyx{\subsection{Block Candidate Selection}}
We experiment the effects of different block candidates on quality, latency, and memory overhead. 
As shown in Table~\ref{tab:block_index}, the more block candidates, the more extra memory overhead will be incurred, as more block outputs need be saved for caching.
However, a larger number of block candidates does not necessarily lead to improvements in generation quality and speed. 
In addition, if there are too few block candidates, the generation quality will drop significantly, since the block level caching cannot capture the different features of different blocks. 
Therefore, we select (10, 20, 30) as the default block configuration, which has the optimal performance in terms of quality and speed in our experiments, and does not incur excessive memory overhead.

\begin{table}[t]
\resizebox{0.48\textwidth}{!}{%
\setlength{\tabcolsep}{1mm}
\begin{tabular}{c|ccc|c|c }
\hline
                 \textbf{Block Candidate}             & \textbf{LPIPS ↓} & \textbf{PSNR ↑}  & \textbf{SSIM ↑} & \textbf{Latency} & \textbf{Mem.} \\ \hline
\textbf{(20)}              & 0.093           & 25.04          & 0.803          & 483 s            &    0.64 GB \\ 
\textbf{(10,30)}            & 0.089           & 25.22          & 0.821          & 470 s           &    1.27 GB     \\ 
\rowcolor{gray!15}\textbf{(10,20,30)} & \textbf{0.079 }       & \textbf{25.91 }         & \textbf{0.858  }        & \textbf{465 s }        &   \textbf{1.89 GB} \\
\textbf{(10,15,20,25,30,35)}   & 0.080           & 25.86       & 0.849          & 468 s            &    3.77 GB     \\ 
\hline
\end{tabular}
}
\caption{Quality, latency and memory overhead under different block candidates on Wan 14B 480p.}
\label{tab:block_index}
\vspace{-15pt}
\end{table}

\vspace{10pt}
\subsection{Scaling to Multi-GPUs and Higher Resolution}
\begin{figure}[h]
\centering
\includegraphics[width=1.0\columnwidth]{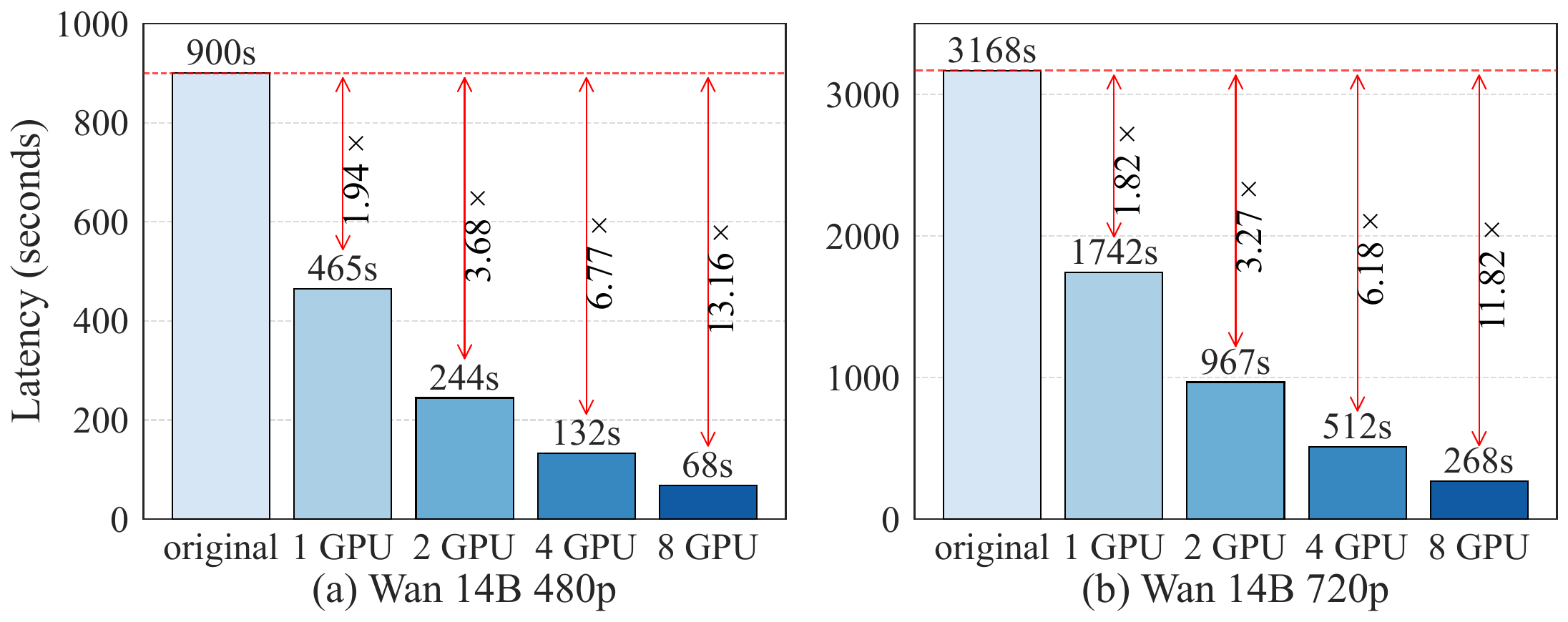}
\caption{Acceleration efficiency of MixCache with different video resolutions and GPU configurations.}
\label{fig:scaling}
\end{figure}

MixCache can be compatible with the current mainstream DiT parallel methods with a minor cache selection synchronization overhead.
We integrate Ulysses parallel~\cite{Ulysses} to MixCache and present the latency on Wan 14B.
As in Figure~\ref{fig:scaling}, the parallel version of MixCache still demonstrates a consistent strong scaling with increasing GPU configurations. 
Besides, this performance advantage is maintained across varying resolutions, demonstrating its effectiveness on high-resolution video generation tasks. 

\section{Conclusion}
To address the high-latency challenge of video DiT inference caused by its multi-step iterative denoising process, we propose MixCache, a training-free caching-based framework for efficient video DiT inference. 
By leveraging the redundancy in the diffusion process of different granularities and adaptively combines three-level caching (step/cfg/block), MixCache achieves comparable generation quality while significantly improving inference efficiency, outperforming existing baselines with a maximum speedup of 1.94$\times$ on Wan 14B and 1.97$\times$ on HunyuanVideo.
\wyxnew{Our work establishes hybrid caching as a novel and effective approach for accelerating video DiT inference, supporting responsive multimedia content generation.} 

\clearpage
\bibliographystyle{ACM-Reference-Format}
\bibliography{ref}

\end{document}